\documentclass[prl,aps,showpacs,twocolumn,floatfix]{revtex4}

\usepackage{graphicx}
\usepackage{amssymb}
\usepackage{latexsym}

\newcommand{\fig}[1]{Fig.~\ref{#1}}

\newcommand{\eq}[1]{Eq.~(\ref{#1})}

\newcommand{\brefs}[1]{Refs.~\cite{#1}}

\newcommand{\ub}{{\rm ub}}
\newcommand{\bd}{{\rm b}}
\newcommand{\piad}{\pi_{\rm ad}}

\begin{document}

\title{Active diffusion of motor particles}
\author{Stefan Klumpp} 
\author{Reinhard Lipowsky} 
\affiliation{Max-Planck-Institut f\"ur Kolloid- und
  Grenzfl\"achenforschung, 14424 Potsdam-Golm, Germany}

\date{\today}

\begin{abstract}
  The movement  of motor particles  consisting of
one or several molecular motors bound to a cargo particle
is studied theoretically. The particles move on
patterns of immobilized  filaments.  Several patterns are
described for which the motor particles undergo non-directed but
enhanced  diffusion. Depending on the walking distance of the
particles and the mesh size of the patterns,  the active
diffusion coefficient  exhibits  three different regimes.
For micrometer-sized motor particles in water,  e.g.,
this diffusion coefficient
can be enhanced by two orders  of magnitude.
\end{abstract}



\maketitle

{\em Introduction.}  Biological motor molecules such as the
cytoskeletal motors kinesin and myosin, which convert the chemical
free energy released from the hydrolysis of adenosine triphosphate
(ATP) into directed movements along cytoskeletal filaments, represent
powerful nanomachines.  It seems rather promising to construct
artificial biomimetic devices based on these molecular motors and to
use them, e.g., for the transport of cargo particles in the nano- and
micrometer regime \cite{Hess_Vogel2001,Boehm_Unger2004}.  These
approaches are based on two different {\it in vitro} motility assays,
the gliding assay, for which molecular motors are anchored at a
substrate surface and drag filaments along this surface, and the bead
assay, for which the filaments are immobilized on the surface and the
motors pull beads or cargo along these filaments \cite{Howard2001}.

In order to obtain {\it spatial control} of the movements, motor
molecules or filaments can be bound to the substrate surface in well
defined patterns by controlling the surface topography and/or
chemistry. Examples are motor-covered surface channels or grooves
which guide the gliding filaments, chemical patterning of a surface to
bind filaments or motors selectively, and combinations of topographic
and chemical patterning, see, e.g.,
\cite{Turner__Murphy1995,Suzuki__Mashiko1997,Riveline__Prost1998,Hess__Vogel2001,Hiratsuka__Uyeda2001,Boehm__Unger2001,Limberis__Stewart2001}.
In this way, one can create 'active stripes' along which the active
transport takes place.  The {\it directionality} of these stripes can
be controlled in a rather direct way for systems based on the bead
assay, since unidirectional movement of motors can be realized by
aligning filaments in parallel with the same orientation
\cite{Boehm__Unger2001,Limberis__Stewart2001}.  In gliding assays,
movement of filaments usually occurs randomly in both directions, but
unidirectional movements can be obtained constructing special surface
domain patterns \cite{Hiratsuka__Uyeda2001}.

Usually, molecular motors 
undergo {\it directed movements} which can be used for transport 
over large distances. 
However, the active motion of
molecular motors can also be used to generate {\it effectively diffusive
movements} by sequences of active directed movements into random
directions. We will call these ATP-dependent and thus energy-consuming,
but effectively non-directed movements \emph{active diffusion}.
It should be useful under conditions
where simple diffusion is slow, i.e., for 
large cargoes or high viscosity of the fluid. 
Active diffusion is also present 
{\it in vivo}, where various motor-driven cargoes such as vesicles, 
messenger RNA, and viruses exhibit active back-and-forth movements 
\cite{Gross2004}.

In the following, we will discuss active diffusion in several simple
systems which can be realized experimentally. We present results 
for several cases where patterns of 'active
stripes' have been created on a surface by immobilizing cytoskeletal
filaments in certain well-defined patterns, so that they 
constitute  a system of tracks 
for the active movements of particles coated with molecular
motors. Even though we focus on these latter systems
derived from bead assays with mobile motors,  our arguments can be easily 
extended to mobile filaments gliding 
in motor-covered surface grooves.

\begin{figure}[b]
  \centering
  \includegraphics[angle=0,width=.9\columnwidth]{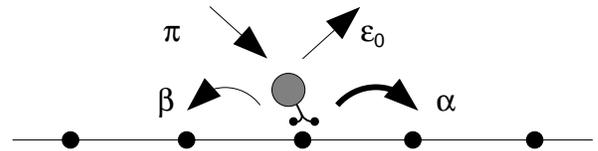}
  \caption{Lattice model for the active motor movements along filaments. 
    Motors make forward and backward steps along the filament with 
    probabilities $\alpha$ and $\beta$, respectively and no step with 
    probability $\gamma$. Unbinding from the filament occurs with 
    probability $\epsilon_0=n_{\rm ad}\epsilon/6$ where $n_{\rm ad}$ is 
    the number of non-filament neighbor sites, and a motor reaching the 
    filament binds to it with  sticking probability $\piad$.}
  \label{fig:model}
\end{figure}

{\em Model.} In order to study the effective diffusion arising from
active motion along patterns of filaments, we describe these movements
as random walks on a lattice
\cite{Lipowsky__Nieuwenhuizen2001,Nieuwenhuizen__Lipowsky2002}. We
consider a cubic lattice with lattice constant $\ell$ on which certain
lines of lattice sites represent the filaments or active stripes.  If
a motor-driven particle occupies such a lattice site, it makes a
forward step (along the orientation of the filament), a backward step,
and no step with probabilities $\alpha$, $\beta$, and $\gamma$ per
time unit $\tau$, respectively.  In addition, it may escape to each of
the $n_{\rm ad}$ adjacent non-filament sites with probability
$\epsilon/6$ ($n_{\rm ad}=3$ and 4 for a filament at a wall and in
solution, respectively).  An unbound particle, i.e., a particle at a
non-filament site, performs a symmetric random walk and steps to each
neighbor site with probability $1/6$ per time $\tau$. An unbound motor
reaching a filament site binds to it with probability $\piad$.  The
parameters of the random walks can be adapted to the measured
transport properties \cite{Lipowsky__Nieuwenhuizen2001}. Confining
surfaces 
are implemented by rejecting all stepping attempts
to lattice sites of the surfaces. 

For the simulations, we chose the basic length scale 
to be $\ell=50$nm and the basic time scale to be $\tau=25$ms. These values
result in a
diffusion coefficient 
$D_\ub\sim 10^{-1}\mu{\rm m}/{\rm s}$ as appropriate for a
micrometer-sized particle in water. We take $\alpha=0.5$, $\beta=0$,
and $\piad=1$, corresponding to a velocity $v_\bd\simeq 1\mu{\rm
  m}/{\rm s}$ and a binding rate 
in the order of $100{\rm s}^{-1}$ for motors within a capture range of
50nm from the filament.

{\em Active diffusion in 1d.} In order to discuss the simplest case of
active diffusion, let us consider two parallel filaments
with opposite polarity. We denote the coordinate axis parallel to
the filaments by $x$ and take the filaments to be sufficiently
long, so that the filament ends are not reached within the observation
time. The filaments are placed within a long
and thin tube or channel with a rectangular cross section (with width 
$L$ and height $H$), so that
unbound motors cannot escape, but rebind with a finite mean return
time, a situation that mimics the axon of a nerve cell.

\begin{figure}[tb]
  \includegraphics[angle=-90,width=.9\columnwidth]{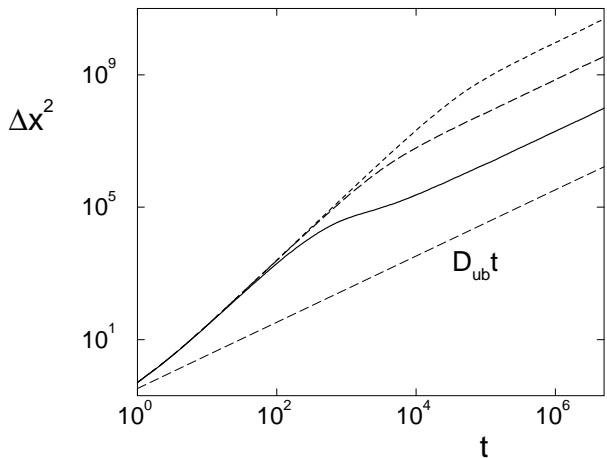}
  \caption{Active diffusion in a tube with two antiparallel filaments 
    (double-logarithmic plot): The 
    variance $\Delta x^2$ of the particle position as a fuction of time 
    $t$ for motor-driven particles  with unbinding probabilities 
    $\epsilon=0.01$ (solid), $10^{-3}$ (dashed), and $10^{-4}$ (dotted 
    line) and other parameters as described in the text. The thin dashed 
    line indicates unbound diffusion which is much smaller than active 
    diffusion. Here and in the following figures, 
    lengths and times are expressed in units of $\ell$ and $\tau$, 
    respectively.}
  \label{fig:actDiff1d}
\end{figure}

We have studied this case using an analytical Fourier--Laplace
technique as in
\brefs{Nieuwenhuizen__Lipowsky2002} and
computer simulations. Some results are shown in \fig{fig:actDiff1d}.
Because of the antiparallel orientation of the filaments, no net
movement is obtained in such a system, and the effective velocity is
zero on large scales.  The effective diffusion coefficient, 
$D\equiv \lim_{t\to \infty}\Delta x^2(t)/2t$, defined via
the long-time behaviour of the positional variance
$\Delta x^2\equiv\langle x^2\rangle -\langle x\rangle^2$,
is found to be given by
\begin{equation}\label{eq:actDiff1d}
  D =  \bar D + D_{\rm act}P_\bd,
\end{equation}
where the first term is the weighted average of the bound and unbound
diffusion coefficient, $\bar D= D_\bd P_\bd+D_\ub (1-P_\bd)$ with the
probability $P_\bd=1/[1+(\epsilon/\piad) L H/\ell^2]$ that the motor 
is bound to
the filament at large times. 
The
contribution of the bound diffusion coefficient is usually small, 
so that $\bar D\simeq D_\ub (1-P_\bd)$. The second
term describes active diffusion, i.e., additional spreading of a
distribution of many motors due to the active directed movements into
both directions with
\begin{equation}
  D_{\rm act}=\frac{3 v_\bd^2}{2\epsilon}.
\end{equation}
The latter contribution can be written as $D_{\rm
  act}=\frac{1}{2}\langle L_s^2\rangle/\Delta t_\bd=\Delta
x_\bd v_\bd$ with the average walking time $\Delta
t_\bd=3/(2\epsilon)$, the average walking distance $\Delta
x_\bd=v_\bd\Delta t_\bd$, and the mean square 
of the
distance $L_s$ traveled actively before a change of direction 
via unbinding, as obtained from the distribution 
$P(L_s)=\frac{1}{2\Delta x_\bd}e^{-|L_s|/\Delta x_\bd}$ of the
walking distances.

This contribution is present even if the diffusive motion of unbound 
motors is strongly suppressed, e.g.\ because of their large cargoes or 
because of a high
viscosity of the fluid. In cells, this should apply to the movements of
large organelles such as mitochondria (where, however, the switching
of direction is accomplished via different types of motors rather than
via the orientation of the filaments) \cite{Gross2004}.

Active diffusion as given by $\Delta x^2\approx 2 D t$ 
is valid at large
times with $t\gg t_*\equiv 2D/v_\bd^2\simeq 3P_\bd/\epsilon$, see
\fig{fig:actDiff1d}.
Note that the larger the active diffusion coefficient, the larger is
also the crossover time $t_*$ and, thus, the time necessary to observe
the active diffusion.

An enhancement of diffusion occurs also in rather different contexts
such as electrophoresis, chromatographic columns, and diffusion in
hydrodynamic flow which can also be described by multi-state random
walk models \cite{vandenBroeck1990,Weiss1994}. In all of these cases,
however, the directed motion arises from externally applied fields
rather than from the active movements of the motor particles.

\begin{figure}[tb]
  \includegraphics[angle=0,width=.95\columnwidth]{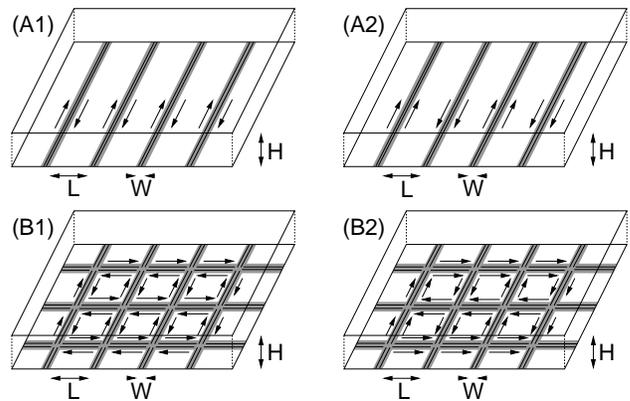}
  \caption{Substrate surfaces with active stripes which consist of filaments 
    immobilized onto striped surface domains: For the system architectures 
    (A1) and (B1), all stripes contain filaments with both orientations, 
    for (A2) 
    and (B2), the filaments within one stripe have the same orientation, 
    but filaments on neighboring stripes have opposite orientation. 
      }
  \label{fig:stripesPatterns}
\end{figure}

{\em Arrays of filaments on a surface. }  
Let us now consider
filaments which are immobilized on patterned surfaces with
two-dimensional arrays of stripes.  We consider four systems as shown
in \fig{fig:stripesPatterns}: (A1, A2) consist of arrays of parallel
stripes whereas (B1, B2) correspond to intersecting stripes which form
a two-dimensional square lattice with lattice constant $L$. The
stripes have a width $W$; in our simulations, we used $W=\ell$.
In both cases, we studied two situations: (i) Each stripe contains
filaments with both possible orientations as in (A1, B1), and motor
particles which bind to a stripe can start to move in either direction
(which is chosen randomly in the simulations); and (ii) neighboring
stripes are covered by filaments with opposite orientation as in (A2,
B2). Note that all patterns have no directional bias on large scales.

\begin{figure}[tb]  
  \includegraphics[angle=-90,width=.95\columnwidth]{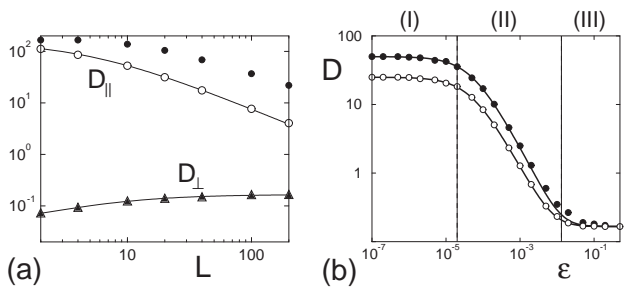}
  \caption{(a) Effective diffusion coefficients $D_\parallel$ (circles) and 
    $D_\perp$ (triangles) as functions of the stripe separation $L$.
    The open and filled symbols correspond to the two system
    architectures (A1) and (A2), respectively, compare
    \fig{fig:stripesPatterns}.  The full lines indicate the analytical
    result for (A1). (b) Effective diffusion coefficient $D$ as a
    function of the unbinding probability $\epsilon$. $D$ exhibits three 
    regimes (I)--(III) as discussed in the text. The open and
    filled symbols correspond to the two system architectures (B1) and
    (B2), respectively, as in \fig{fig:stripesPatterns}. The
    parameters are as described in the text; we used
    $\epsilon=0.0025$, $H=100\ell$ in (a) and $L=H=100\ell$ in (b). }
  \label{fig:Deff}
\end{figure}

The motors can change their direction of movement via unbinding and
rebinding to another filament. In the stripe lattices, they can also
change direction at the vertices by switching to another filament,
which is chosen randomly. Finally, we take the diffusion in the
direction perpendicular to the surface to be restricted to a slab of
height $H$.  To study the active diffusion for this case, we
determined the diffusion coefficient by extensive Monte Carlo
simulations for motor-driven active particles of micrometer size and
for a wide range of the unbinding rate $\epsilon$ and the stripe
separation $L$ as shown in \fig{fig:Deff}.

Arrangements of filaments of this type are accessible to experiments
and can be created using structured surfaces as described above.
Lattice patterns of filaments can also be formed by the
self-organization of filaments and motor complexes
\cite{Surrey__Karsenti2001} or by the assembly of crosslinked networks
on micropillars \cite{Roos__Spatz2003}.

{\em Parallel active stripes. } 
Arrays of parallel active stripes which contain filaments with both
orientations (A1) exhibit active diffusion very similar to the
one-dimensional case discussed above.  However, in this case, the
effective diffusion coefficient depends on the spatial direction.
While the effective diffusion is described by \eq{eq:actDiff1d} in the
direction parallel to the stripes, which leads to the parallel diffusion 
coefficient $D_\parallel=\bar D+D_{\rm
  act}P_\bd$, the diffusion perpendicular to the stripes is
characterized by $D_\perp=D_\ub(1-P_\bd)$, see \fig{fig:Deff}(a).  The
latter expression is also valid if the directionality of the stripes
alternates (A2), where the unbound motors return predominantly to the
same stripe which leads to a higher value of $D_\parallel$ as shown in
\fig{fig:Deff}(a).

{\em  Regular lattices of active stripes. }
In the system architectures (B1,B2), the motor-driven
active particles can perform two-dimensional random walks on the
stripe lattice even if they remain bound to the filaments. In addition,
effectively diffusive behavior can arise from the repeated unbinding
from and binding to filament. Let us consider case (B1).  The
diffusion coefficient is given by $D=D_{\rm act} P_\bd+D_\ub
(1-P_\bd)$ with $P_\bd
\approx
1/[1+(\epsilon/\piad)LH/(2 W \ell)]$ for the stripe lattice. Depending on the
processivity of the active particles, i.e.\ on the value of the
unbinding rate, there are three regimes:

(I) If the unbinding rate is small, so that the walking distance is large 
compared to the pattern mesh size, $\Delta x_\bd=2v_\bd/\epsilon\gg L$,
the particle performs a 
random
walk along the active stripe lattice
and the diffusion coefficient is
$D\approx D(\epsilon=0)=\frac{1}{4}v_\bd L$.
This case should be appropriate for cargo particles driven by many
processive motors.

(II) If the average walking distance before unbinding is smaller than
the mesh size, 
$\Delta x_\bd=2v_\bd/\epsilon\ll L$ as
appropriate, e.g., for transport by single processive motors, changes
in direction of movement will mainly be through unbinding and
rebinding to another filament. In that case, the distance traveled 
before a change of direction is given by the walking distance
$\Delta x_\bd$ 
and, similar to the
one-dimensional case discussed before, the diffusion coefficient is
$D\approx \frac{1}{2}\Delta x_\bd v_\bd P_\bd$ \cite{endn2}.

(III) Finally, if motor particles spend most of the time unbound, 
i.e., for large $\epsilon$ as appropriate for weakly processive
motors, the diffusion coefficient is simply given by the unbound
Brownian motion, $D\approx D_\ub$. 

The three regimes can be summarized in one formula by using $D_{\rm
  act}=v_\bd L_s/4$ and defining the interpolated 
distance traveled between two changes of direction by
$L_s=L 2\Delta x_\bd/(L+2\Delta x_\bd)$ which leads to
\begin{equation}\label{deff_formula}
  D=\frac{v_\bd}{2}\frac{L\Delta x_\bd}{L+2\Delta x_\bd} P_\bd+D_\ub (1-P_\bd).
\end{equation}
The latter expression leads to the solid line in
\fig{fig:Deff}(b), which agrees well with the data points
(open circles) from our simulations.

The value of the active diffusion coefficient can be further enhanced
by controlling the orientation of the filaments along the stripes,
i.e., by using filament patterns such as (B2) for which 
we obtain an active diffusion coefficient twice as
large as for case (B1). In regime (I) this factor is due to the fact
that a fully random choice of the new direction is completed after a
walk over the distance $2L$. However, \eq{deff_formula} with an
additional factor of two in the active diffusion term describes rather
well our simulation data also in regime (II).

{\em  Discussion. } Finally, let us estimate the order of magnitude of
the active diffusion in systems that are experimentally accessible.
Free diffusion in a Newtonian fluid is governed by the Stokes--Einstein 
relation
from which the diffusion coefficient is obtained as $D_\ub\simeq (100
{\rm nm}/R_{\rm hyd})\times 2.4 \mu{\rm m}^2/{\rm s} \times \eta_{\rm
  water}/\eta$. In the latter expression, $R_{\rm hyd}$ is the
effective hydrodynamic radius of the particle and $\eta$ the 
viscosity
of the medium. For a micrometer-sized particle in water, we have
$D_\ub\sim 0.1\mu{\rm m}^2/$s, for a 100nm-sized particle $D_\ub\sim
1\mu{\rm m}^2/$s. 
(The diffusion coefficient of bound motors moving along filaments is much 
smaller, typically of the order of 
$D_\bd\sim v_\bd\ell\sim 0.01 \mu{\rm m}^2/$s
with $v_\bd\sim
1\mu$m$/$s and $\ell\sim 10$nm.)

The active diffusion coefficient, on the other hand, is
of the order of
$\frac{1}{4}v_\bd L_s P_\bd$ with $L_s$ given by the filament length
in regime (I) and by the walking distance in regime (II). With a
typical filament length of a few tens of $\mu$m, this means that
active diffusion coefficients of the order of 10$\mu{\rm m}^2/$s can
be obtained for cargo particles driven by many motors, for which
regime (I) is appropriate. For micrometer-sized particles this is two
orders of magnitude larger than the diffusion coefficient of unbound
Brownian motion.
For cargo particles driven by a single or a few
motors, regime (II) applies, and for a typical unbinding rate of 
$1/$s, we obtain an active diffusion coefficient of $\sim
1\mu{\rm m}^2/$s. 

These estimates as well as our simulation data are for diffusion in
water. In a more viscous medium with viscosity $\eta>\eta_{\rm
  water}$, diffusion is even more enhanced due
to the active processes, because active diffusion does not obey the 
Einstein relation and is largely independent of $\eta$ \cite{endn1}: 
the viscous force on a micrometer-sized bead moving with $1\mu{\rm
  m}/{\rm s}$ is of the order of $\sim \eta/\eta_{\rm water}\times
10^{-2}\,{\rm pN}$ which is small compared to the piconewton forces
required to slow down the active movements of a single motor. Unbound
diffusion is however reduced by the factor $\eta_{\rm water}/\eta$, so
that the ratio $D/D_\ub\sim\eta$ increases with increasing viscosity
$\eta$.

Active diffusion will therefore be useful for particles which explore
surface patterns with linear dimensions of tens of micrometers or even
millimeters, e.g., in order to find a specific binding partner or to
deliver a cargo, in particular if the surrounding fluid has a high
viscosity.  One potential application is provided by the integration
of these biomimetic transport systems into labs-on-a-chip for genomics
or proteomics.

In summary, we propose here to use chemically or topographically
patterned surfaces to create well defined arrangements of active
stripes covered by filaments and to use the random walks arising from
the active movements of motor-covered particles along these stripes to
achieve fast diffusion on the surface.


\begin{thebibliography}{18}
\expandafter\ifx\csname natexlab\endcsname\relax\def\natexlab#1{#1}\fi
\expandafter\ifx\csname bibnamefont\endcsname\relax
  \def\bibnamefont#1{#1}\fi
\expandafter\ifx\csname bibfnamefont\endcsname\relax
  \def\bibfnamefont#1{#1}\fi
\expandafter\ifx\csname citenamefont\endcsname\relax
  \def\citenamefont#1{#1}\fi
\expandafter\ifx\csname url\endcsname\relax
  \def\url#1{\texttt{#1}}\fi
\expandafter\ifx\csname urlprefix\endcsname\relax\def\urlprefix{URL }\fi
\providecommand{\bibinfo}[2]{#2}
\providecommand{\eprint}[2][]{\url{#2}}

\bibitem[{\citenamefont{Hess and Vogel}(2001)}]{Hess_Vogel2001}
\bibinfo{author}{\bibfnamefont{H.}~\bibnamefont{Hess}} \bibnamefont{and}
  \bibinfo{author}{\bibfnamefont{V.}~\bibnamefont{Vogel}},
  \bibinfo{journal}{Rev.\ Mol.\ Biotechnology} \textbf{\bibinfo{volume}{82}},
  \bibinfo{pages}{67} (\bibinfo{year}{2001}).

\bibitem[{\citenamefont{{B\"ohm} and Unger}(2004)}]{Boehm_Unger2004}
\bibinfo{author}{\bibfnamefont{K.~J.} \bibnamefont{{B\"ohm}}} \bibnamefont{and}
  \bibinfo{author}{\bibfnamefont{E.}~\bibnamefont{Unger}}, in
  \emph{\bibinfo{booktitle}{Encyclopedia of Nanoscience and Nanotechnology}},
  edited by \bibinfo{editor}{\bibfnamefont{H.~S.} \bibnamefont{Nalwa}}
  (\bibinfo{publisher}{American Scientific Publishers},
  \bibinfo{address}{Stevenson Ranch}, \bibinfo{year}{2004}),
  vol.~\bibinfo{volume}{4}, pp. \bibinfo{pages}{345--357}.

\bibitem[{\citenamefont{Howard}(2001)}]{Howard2001}
\bibinfo{author}{\bibfnamefont{J.}~\bibnamefont{Howard}},
  \emph{\bibinfo{title}{Mechanics of Motor Proteins and the Cytoskeleton}}
  (\bibinfo{publisher}{Sinauer Associates}, \bibinfo{address}{Sunderland}, 
  \bibinfo{year}{2001}).

\bibitem[{\citenamefont{Suzuki et~al.}(1997)\citenamefont{Suzuki, Yamada, Oiwa,
  Nakayama, and Mashiko}}]{Suzuki__Mashiko1997}
\bibinfo{author}{\bibfnamefont{H.}~\bibnamefont{Suzuki}},
  \bibinfo{author}{\bibfnamefont{A.}~\bibnamefont{Yamada}},
  \bibinfo{author}{\bibfnamefont{K.}~\bibnamefont{Oiwa}},
  \bibinfo{author}{\bibfnamefont{H.}~\bibnamefont{Nakayama}}, \bibnamefont{and}
  \bibinfo{author}{\bibfnamefont{S.}~\bibnamefont{Mashiko}},
  \bibinfo{journal}{Biophys.\ J.} \textbf{\bibinfo{volume}{72}},
  \bibinfo{pages}{1997} (\bibinfo{year}{1997}).

\bibitem[{\citenamefont{Riveline et~al.}(1998)\citenamefont{Riveline, Ott,
  J{\"u}licher, Winkelmann, Cardoso, Lacap{\`e}re, Magn{\'u}sd{\'o}ttir, Viovy,
  Gorre-Talini, and Prost}}]{Riveline__Prost1998}
\bibinfo{author}{\bibfnamefont{D.}~\bibnamefont{Riveline}},
  \bibinfo{author}{\bibfnamefont{A.}~\bibnamefont{Ott}},
  \bibinfo{author}{\bibfnamefont{F.}~\bibnamefont{J{\"u}licher}},
  \bibinfo{author}{\bibfnamefont{D.~A.} \bibnamefont{Winkelmann}},
  \bibinfo{author}{\bibfnamefont{O.}~\bibnamefont{Cardoso}},
  \bibinfo{author}{\bibfnamefont{J.-J.} \bibnamefont{Lacap{\`e}re}},
  \bibinfo{author}{\bibfnamefont{S.}~\bibnamefont{Magn{\'u}sd{\'o}ttir}},
  \bibinfo{author}{\bibfnamefont{J.-L.} \bibnamefont{Viovy}},
  \bibinfo{author}{\bibfnamefont{L.}~\bibnamefont{Gorre-Talini}},
  \bibnamefont{and} \bibinfo{author}{\bibfnamefont{J.}~\bibnamefont{Prost}},
  \bibinfo{journal}{Eur.\ Biophys.\ J.} \textbf{\bibinfo{volume}{27}},
  \bibinfo{pages}{403} (\bibinfo{year}{1998}).

\bibitem[{\citenamefont{Hess et~al.}(2001)\citenamefont{Hess, Clemmens, Qin,
  Howard, and Vogel}}]{Hess__Vogel2001}
\bibinfo{author}{\bibfnamefont{H.}~\bibnamefont{Hess}},
  \bibinfo{author}{\bibfnamefont{J.}~\bibnamefont{Clemmens}},
  \bibinfo{author}{\bibfnamefont{D.}~\bibnamefont{Qin}},
  \bibinfo{author}{\bibfnamefont{J.}~\bibnamefont{Howard}}, \bibnamefont{and}
  \bibinfo{author}{\bibfnamefont{V.}~\bibnamefont{Vogel}},
  \bibinfo{journal}{Nano Lett.} \textbf{\bibinfo{volume}{1}},
  \bibinfo{pages}{235} (\bibinfo{year}{2001}).

\bibitem[{\citenamefont{Hiratsuka et~al.}(2001)\citenamefont{Hiratsuka, Tada,
  Oiwa, Kanayama, and Uyeda}}]{Hiratsuka__Uyeda2001}
\bibinfo{author}{\bibfnamefont{Y.}~\bibnamefont{Hiratsuka}},
  \bibinfo{author}{\bibfnamefont{T.}~\bibnamefont{Tada}},
  \bibinfo{author}{\bibfnamefont{K.}~\bibnamefont{Oiwa}},
  \bibinfo{author}{\bibfnamefont{T.}~\bibnamefont{Kanayama}}, \bibnamefont{and}
  \bibinfo{author}{\bibfnamefont{T.~Q.~P.} \bibnamefont{Uyeda}},
  \bibinfo{journal}{Biophys.\ J.} \textbf{\bibinfo{volume}{81}},
  \bibinfo{pages}{1555} (\bibinfo{year}{2001}).

\bibitem[{\citenamefont{Turner et~al.}(1995)\citenamefont{Turner, Chang, Fang,
  Brandow, and Murphy}}]{Turner__Murphy1995}
\bibinfo{author}{\bibfnamefont{D.~C.} \bibnamefont{Turner}},
  \bibinfo{author}{\bibfnamefont{C.}~\bibnamefont{Chang}},
  \bibinfo{author}{\bibfnamefont{K.}~\bibnamefont{Fang}},
  \bibinfo{author}{\bibfnamefont{S.~L.} \bibnamefont{Brandow}},
  \bibnamefont{and} \bibinfo{author}{\bibfnamefont{D.~B.}
  \bibnamefont{Murphy}}, \bibinfo{journal}{Biophys.\ J.}
  \textbf{\bibinfo{volume}{69}}, \bibinfo{pages}{2782} (\bibinfo{year}{1995}).

\bibitem[{\citenamefont{B{\"o}hm et~al.}(2001)\citenamefont{B{\"o}hm, Stracke,
  M{\"u}hlig, and Unger}}]{Boehm__Unger2001}
\bibinfo{author}{\bibfnamefont{K.~J.} \bibnamefont{B{\"o}hm}},
  \bibinfo{author}{\bibfnamefont{R.}~\bibnamefont{Stracke}},
  \bibinfo{author}{\bibfnamefont{P.}~\bibnamefont{M{\"u}hlig}},
  \bibnamefont{and} \bibinfo{author}{\bibfnamefont{E.}~\bibnamefont{Unger}},
  \bibinfo{journal}{Nanotechnology} \textbf{\bibinfo{volume}{12}},
  \bibinfo{pages}{238} (\bibinfo{year}{2001}).

\bibitem[{\citenamefont{Limberis et~al.}(2001)\citenamefont{Limberis, Magda,
  and Stewart}}]{Limberis__Stewart2001}
\bibinfo{author}{\bibfnamefont{L.}~\bibnamefont{Limberis}},
  \bibinfo{author}{\bibfnamefont{J.~J.} \bibnamefont{Magda}}, \bibnamefont{and}
  \bibinfo{author}{\bibfnamefont{R.~J.} \bibnamefont{Stewart}},
  \bibinfo{journal}{Nano Lett.} \textbf{\bibinfo{volume}{1}},
  \bibinfo{pages}{277} (\bibinfo{year}{2001}).

\bibitem[{\citenamefont{Gross}(2004)}]{Gross2004}
\bibinfo{author}{\bibfnamefont{S.~P.} \bibnamefont{Gross}},
  \bibinfo{journal}{Phys. Biol.} \textbf{\bibinfo{volume}{1}},
  \bibinfo{pages}{R1} (\bibinfo{year}{2004}).

\bibitem[{\citenamefont{Lipowsky et~al.}(2001)\citenamefont{Lipowsky, Klumpp,
  and Nieuwenhuizen}}]{Lipowsky__Nieuwenhuizen2001}
\bibinfo{author}{\bibfnamefont{R.}~\bibnamefont{Lipowsky}},
  \bibinfo{author}{\bibfnamefont{S.}~\bibnamefont{Klumpp}}, \bibnamefont{and}
  \bibinfo{author}{\bibfnamefont{T.~M.} \bibnamefont{Nieuwenhuizen}},
  \bibinfo{journal}{Phys.\ Rev.\ Lett.} \textbf{\bibinfo{volume}{87}},
  \bibinfo{pages}{108101} (\bibinfo{year}{2001}).

\bibitem[{\citenamefont{Nieuwenhuizen et~al.}(2002)\citenamefont{Nieuwenhuizen,
  Klumpp, and Lipowsky}}]{Nieuwenhuizen__Lipowsky2002}
\bibinfo{author}{\bibfnamefont{T.~M.} \bibnamefont{Nieuwenhuizen}},
  \bibinfo{author}{\bibfnamefont{S.}~\bibnamefont{Klumpp}}, \bibnamefont{and}
  \bibinfo{author}{\bibfnamefont{R.}~\bibnamefont{Lipowsky}},
  \bibinfo{journal}{Europhys.\ Lett.} \textbf{\bibinfo{volume}{58}},
  \bibinfo{pages}{468} (\bibinfo{year}{2002}) and \bibinfo{journal}{Phys.\ Rev.\ E} \textbf{\bibinfo{volume}{69}},
  \bibinfo{pages}{061911} (\bibinfo{year}{2004}).

\bibitem[{\citenamefont{Weiss}(1994)}]{Weiss1994}
\bibinfo{author}{\bibfnamefont{G.}~\bibnamefont{Weiss}},
  \emph{\bibinfo{title}{Aspects and Applications of the Random Walk}}
  (\bibinfo{publisher}{North-Holland}, \bibinfo{address}{Amsterdam},
  \bibinfo{year}{1994}).

\bibitem[{\citenamefont{{Van den Broeck}}(1990)}]{vandenBroeck1990}
\bibinfo{author}{\bibfnamefont{C.}~\bibnamefont{{Van den Broeck}}},
  \bibinfo{journal}{Physica A} \textbf{\bibinfo{volume}{168}},
  \bibinfo{pages}{677} (\bibinfo{year}{1990}).

\bibitem[{\citenamefont{Surrey et~al.}(2001)\citenamefont{Surrey,
  {N\'ed\'elec}, Leibler, and Karsenti}}]{Surrey__Karsenti2001}
\bibinfo{author}{\bibfnamefont{T.}~\bibnamefont{Surrey}},
  \bibinfo{author}{\bibfnamefont{F.}~\bibnamefont{{N\'ed\'elec}}},
  \bibinfo{author}{\bibfnamefont{S.}~\bibnamefont{Leibler}}, \bibnamefont{and}
  \bibinfo{author}{\bibfnamefont{E.}~\bibnamefont{Karsenti}},
  \bibinfo{journal}{Science} \textbf{\bibinfo{volume}{292}},
  \bibinfo{pages}{1167} (\bibinfo{year}{2001}).

\bibitem[{\citenamefont{Roos et~al.}(2003)\citenamefont{Roos, Roth, Konle,
  Presting, Sackmann, and Spatz}}]{Roos__Spatz2003}
\bibinfo{author}{\bibfnamefont{W.~H.} \bibnamefont{Roos}},
  \bibinfo{author}{\bibfnamefont{A.}~\bibnamefont{Roth}},
  \bibinfo{author}{\bibfnamefont{J.}~\bibnamefont{Konle}},
  \bibinfo{author}{\bibfnamefont{H.}~\bibnamefont{Presting}},
  \bibinfo{author}{\bibfnamefont{E.}~\bibnamefont{Sackmann}}, \bibnamefont{and}
  \bibinfo{author}{\bibfnamefont{J.~P.} \bibnamefont{Spatz}},
  \bibinfo{journal}{ChemPhysChem} \textbf{\bibinfo{volume}{4}},
  \bibinfo{pages}{872} (\bibinfo{year}{2003}).

\bibitem{endn2}
The width of regime (II) can be increased by an increase of $v_\bd$ which shifts regime (I) to smaller $\epsilon$ and by increasing $\piad$ which shifts regime (III) to larger $\epsilon$.



\bibitem{endn1}
For a linear force-velocity relation, $v_\bd=v_{\rm b,max}/[1+ 10^{-3}\times \eta/\eta_{\rm water} (R_{\rm hyd}/1 \mu{\rm m})/N]$, where $N$ is the number of motors driving a cargo particle.

\end{thebibliography}
\end{document}